\documentclass[vecphys]{svmult}

\usepackage{makeidx}         
\usepackage{graphicx}        
\usepackage{multicol}        
\usepackage[bottom]{footmisc}
\usepackage{natbib}

\makeindex             

\newcommand{\msc}[1]{{\mbox{\scriptsize{#1}}}}

\begin{document}

\title*{Dynamics and planet formation in/around binaries}
\author{F. Marzari\inst{1}\, P. Th\'ebault\inst{2,3}\, 
S. Kortenkamp\inst{4} \and H. Scholl\inst{5}}
\institute{Dipartimento di Fisica, Via Marzolo 8, 35131 Padova, Italy
\texttt{marzari@pd.infn.it}
\and Observatoire de Paris, Section de Meudon, F-92195 Meudon Principal Cedex,
France \texttt{philippe.thebault@obspm.fr}
\and Stockholm Observatory, Albanova Universitetcentrum,
SE-10691 Stockholm, Sweden \texttt{thebault@astro.su.se}
\and Planetary Science Institute, Tucson, Arizona \texttt{kortenka@psi.edu}
\and Observatoire de la C\^ote d'Azur, Nice, France
\texttt{scholl@obs-nice.fr}
}
%
%
%
\maketitle
\begin{abstract}
\noindent

We study to which extent planetesimal accretion is affected by the perturbing
presence of a compagnon star. We concentrate on one crucial parameter:
the distribution of encounter velocities within the planetesimal swarm.
We numerically explore the evolution of this parameter taking into
account the secular perturbations of the binary and friction due to
the very likely presence of gas in the disk. $<\Delta v>$ maps are derived, for
planetesimals of different sizes, for a total of 120 binary
configurations (eccentricity eb and separation ab).
We identify for each case 3 different accreting behaviours. 1) In regions
where no significant dV increase is observed, ``classical'' (i.e., single-star)
accretion is possible. 2) In regions where $dV>v_{ero}$, the threshold
velocity above which all impacts are eroding, no accretion is possible
and planet growth is stopped. 3) In between these 2 limiting behaviours, a
large fraction of binary configurations leads to significant dV increase, but
still below the erosion threshold. In this intermediate case, planetesimal
growth can occur, but proceeds slower than in the single-star case,
possibly following the so-called type II runaway groth mode.

\end{abstract}
\section{Introduction} 
\label{sec:1}

\subsection{the ``standard'' planet formation scenario}

A fundamental step in the ''standard'' scenario for the formation of planets, 
either terrestrial or gaseous
giant, is the collisional accumulation of  
kilometer--sized planetesimals in successively larger bodies.
It is widely admitted that planetesimals emerge from the 
coagulation of silicate and ice grains in the midplane
of protoplanetary disks on timescales of the order of $10^4$ yrs,
even if the details of the formation process are not well understood 
at present. Colliding with each other, planetesimals grow into 
large planetary embryos (Lunar-- to Mars--size) through
a phase of ``runaway'' and ``oligarchic'' growth 
on a timescale of the order
of $10^{4} - 10^5$ years 
\citep{gre78,wet89,bar93,lis93,kok98,kok00,raf03,raf04}.  This is a robust finding
corroborated by distinct numerical methods \cite[see review by][]{kort2000_oem}.
For terrestrial planets, the formation process reaches completion after
a final phase of giant impacts and mutual accretion 
of planetary embryos 
resulting in $10^{27} - 10^{28}$ gram full--size bodies in about 
$10^7-10^8$ years on stable separated orbits \citep{chwe98}.
In the case of giant planets, 
protoplanets accumulate into a
solid core that, once sufficiently massive, begins to accrete a gaseous
envelope.
Initially, this envelope is in hydrostatic equilibrium sustained by
the luminosity provided by the accreting planetesimals and
is increasing at a slow rate compared
to that of the solid core.
However, at a critical core mass (in between 5 to 15 $M_{\oplus}$)
a phase of runaway gas accretion occurs, the planet mass increases rapidly and 
it reaches its final value on a timescale estimated to be only a few 
$10^3$ yr. For a given density of solids and gas in the disk, 
hydrodynamical calculations can determine for which mass of the 
core the envelope is able to remain static or collapse, and the rate
at which the mass flows onto the planet \cite[see e.g.][]{wupp}.
At the end of the gas rapid infall,
the planet evolves into the isolated stage during which it
cannot grow any further because of the exaustion of the gas supply either 
because of its removal by effects of the star or as a result of 
gap formation around the planet orbit triggered by disk-planet 
tidal interaction \citep{pali}. 
The planet finally  cools down and contracts to its present size. 
This scenario for giant planet formation, called core accretion 
\citep{pollo}, 
is based on the interior models for
Jupiter and Saturn which strongly suggest the presence of a dense central
core surrounded by an envelope composed mainly by hydrogen and helium
for both planets. There are large uncertainties on the mass
of the core and detailed calculations performed by 
\cite[e.g.][]{gui1}
predict un upper limit
of 10 $M_{\oplus}$ for Jupiter's core while for Saturn the 
core mass ranges from 6 to 17 $M_{\oplus}$. 

It is believed that extrasolar planets, detected in orbit around an increasing
number of nearby stars may have formed either by core accretion 
or by disk instability \cite[e.g.][]{bos1}. 
While core accretion requires a few million 
years to form a gas giant planet, in the competing model
self-gravitating clumps formed by disk instability will
contract to planetary densities in times of a few hundred years.
The two formation mechanisms are both viable and, at the moment, there is
no reason to definitively exclude one in favor of the other. In this paper,
however, we will concentrate on planetesimal accretion as a common 
mechanism for forming either terrestrial planets or the core of 
gaseous giants.

In a given protoplanetary disk, 
the potentiality of a planetesimal swarm to accumulate into larger 
bodies is measured by two critical parameters.
\begin{itemize} 
\item {\it The surface density of solid material}. 
The concentration of condensible material determines the final mass in the 
case of a terrestrial planet, while it must lead to a reasonable formation 
time for the core of a giant planet that
must be compatible with the lifetime of the gaseous disk. 
Disk ages are estimated to range from 0.1 to 10 Myr 
\citep{stro,hai,chka},
while 3 Myr is the age at which half of the stars show
evidence of disks. 
Simulations of the accumulation of jovian planets 
\citep{boli}  
show that  
if the surface density of solids in the disk is assumed to be at least 
3--times larger than that of the minimum--mass solar nebula 
\citep{wei77}
then 
the formation time for Jupiter mass planets is within 2--3 Myr 
at 5 AU from the star. If planetary migration is included in 
the simulations, the timescale for forming Jupiter is reduced
to  1­-2 Myr even with solid surface density closer to the
minimum mass solar nebula \citep{riar}. 

\item {\it The relative encounter velocities between planetesimals}.

The random encounter velocities in a planetesimal swarm determine
whether mutual impacts will result in accretion or on the contrary
in cratering or fragmentation. 
The key mechanisms controlling the impact velocities
are:  1) mutual gravitational encounters between 
planetesimals, 2) physical collisions, 3) 
gas drag, in presence of the gaseous component of the disk.
Several numerical studies have shown that 
in an unperturbed swarm of kilometer--sized planetesimals around a single star
the average random encounter velocity $\langle \Delta v \rangle$ is 
low enough to favor the collisional accumulation of bodies \citep[e.g.][]{gre78}.
At the early stages of accretion a low $\langle \Delta v \rangle$ is
also a fundamental requirement for permitting a phase of 
rapid "runaway" growth which significantly 
shortens the timescale for the formation of planetary embryos. 
Indeed, under this condition the ratio $ v_{esc(R)}/\langle \Delta v \rangle$ 
between the escape speed of the largest bodies and the average
impact velocities is $>>1$, thus making their
gravitational cross--section quite larger than the geometrical cross--section.
This gravitational focusing enhances the growth rate of the larger planetesimals
which grow much faster than the rest of the population \citep[e.g.][]{lis93}.
A direct consequence of this crucial $\langle \Delta v \rangle$ dependence is that
planetesimal growth is very sensitive to any increase of
the encounter velocities. It could thus be significantly slowed 
down or even stopped if the gravitational influence of a massive body external
to the swarm, like a a pre-existing giant planets (e.g., Jupiter in our solar system)
or a companion star in a binary system is able to excite large impact velocities.
\cite{the02} have shown that the formation of
terrestrial planets in those exoplanetary systems hosting a giant planet
on an external orbit may have been indeed halted.
They find that in both the $\epsilon$ Eridani and 47 Uma planetary
systems, the relative velocity between planetesimals
in the terrestrial planet region may have
quickly exceed the critical threshold value 
$v_{ero}$ beyond which erosion 
dominates before the completion of the runaway growth
if the massive outer planet rapidly reached its present mass.
Of course, the threshold velocity that preferentially result
in disruption rather than accretion 
is a sensitive
function of the internal strength of planetesimals, their bulk
density and the fraction of impact energy partitioned into the
fragments after the impact. These physical parameters strongly
depend on the initial radial density and temperature profile
in the disk midplane.
\end{itemize} 

\subsection{planets in binaries, planet formation in binaries}

In this paper we will focus on how the presence of a companion star 
affects the planetesimal accretion process either in the formation of 
terrestrial planets or the core of a giant planet. While the 
existence of extrasolar Earth--like planets has yet to be assessed, 
some of the giant exoplanets (up to about 40 so far) have been detected 
in the so--called S--type orbits that encircle one component of a 
binary system \citep[e.g.][]{ragha06,desi07}.
Most of these are widely separated pairs where
the stellar companion probably does not have a significant influence 
on the planetesimal accretion around the other star. 
However, in three of the systems, $\gamma Cephei$ GL 86, and HD41004,
the companion is within 20--25 AU from the star hosting the planet.
The $\gamma Cephei$ system is
made of a central star with an estimated mass of 
$1.40 \pm 0.12 M_{\odot}$, a planet 
orbiting at $2.13 \pm 0.05$ AU with a minimum mass of $1.6 \pm 0.13$ 
$M_J$ and a companion star on an eccentric orbit ($e=0.411 \pm 0.0063$)
with semimajor axis of $20.23 \pm 0.64$ AU and mass of 
about $0.4 M_{\odot}$ \citep{hazz,torr07,neuh07}.
GL 86 is a dwarf star 
somewhat less massive than the Sun \citep[about
$0.7 M_{\odot}$, e.g.][]{sant} with a white dwarf companion
of $0.55 M_{\odot}$ \citep{mune}
on an orbit having a semimajor axis of 18.42 AU and an eccentricity of
e=0.3974  \citep{laga}. A massive jovian 
planet ($M sin (i) = 3.91 M_J$) moves around the primary star on an 
almost circular orbit ($e=0.0416$) with a semimajor axis of $a=0.113$ AU
\citep{egg03}.
HD41004 A is a $0.7 M_{\odot}$ star with a dwarf M4V companion
orbiting at $\simeq 22\,AU$. The primary star has a $M sin (i) = 2.3 M_J$
planetary companion on a very eccentric e=0.39 orbit at
$a=1.31\,$AU \citep{zuck04,ragha06}. Note that this system is a hierarchical
quadruple, since the stellar companion is itself orbited at $a=0.016\,$ AU
by a brown dwarf of minimum mass $18.4 M_J$ \citep{zuck04}.

The observed characteristics of these systems, e.g., the proximity of 
the companion star and its eccentric orbit, strongly suggest that 
the presence of the secondary star must have had an influence
on the formation of the detected planets. 
The gravitational pull of the companion star
acts on the $\langle \Delta v \rangle$ of the planetesimal 
disk around the primary star. As previously mentioned, this
can modify the course of the
accretion process which strongly depends on the 
ratio $v_{esc}/\langle \Delta v \rangle$, where $v_{esc}$ is
the surface escape velocity of the biggest accreting bodies.
Timing is an important issue since the perturbations will 
be effective in altering the planetesimal formation
process only if $t_{sec}$, the timescale 
required by the the secular perturbations
of the star to induce a significant $\Delta v$ increase,
is shorter than 
$t_{grow}$, the timescale for the runaway/oligarchic formation of embryos
which is typically of the order of $10^{4}$ to $10^{5}$yrs.
We can outline  three possible different evolutions
for a perturbed planetesimal swarm: 

\begin{itemize}
\item
1) If $\langle \Delta v \rangle$ remains smaller
than $v_{esc}$ during the growth phase then
Planetary formation proceeds almost unaffected by the companion
perturbations . The ``classical'' planetesimal
accumulation scenario holds.  
\item
2) For values of  $\langle \Delta v \rangle$ larger than 
$ v_{esc}$ but still smaller than $v_{ero}$, the threshold
velocity for erosion to dominate over accretion, 
planetary accretion will still be possible but runaway growth will either 
not occur or will occur only after large enough bodies have formed 
such that $ v_{esc}>\langle \Delta v \rangle$ 
(the so-called Type II runaway growth described in Section 6 below).
\item
3) If $\langle \Delta v \rangle$ is increased 
beyond $v_{ero}$, in the majority of collisions cratering
and fragmentation will overcome accretion and the planetesimal
population will be slowly ground down to dust, failing to
form a full size planet. 
\end{itemize}

In the following, we will focus on the determination of 
$\langle \Delta v \rangle$ for different parameters of the 
binary star system with the goal of evaluating when planetesimal accretion is 
possible. We have performed a 
series of deterministic numerical
simulations, following the dynamical evolution of
a swarm of test planetesimals, under the influence of the
companion star's gravitational pull and friction by the gas
in which the planetesimals are imbedded.
Crucial parameters, such as the companion star's mass $m_b$,
semi--major axis $a_b$ and orbital eccentricity $e_b$ are
explored as free parameters. 
We use a code initially developed for the study of planetesimal
systems perturbed by giant planets \citep{thebra98,thebeu01,the02}
and adapted to the circumprimary case \citep{mascho00,the04,the06}.
This 3-D code has a close encounter tracking algorithm which
enables precise determination of the $\Delta v$ within the system.

In a first step we present a detailed study of 
the planetesimal secular dynamics in presence of a companion star (section 2).
We will then proceed by introducing the effects of gas drag and 
by estimating how the combined action of the friction force with 
the secular perturbations affect the 
value of $\langle \Delta v \rangle$ (section 3). We then compare 
the derived $\langle \Delta v \rangle$ with an an approximate estimate 
of $v_{ero}$ to derive a yes--or--no criterion for planetary formation 
as a function of the binary orbital parameters (section 4). 
Finally, in section 5, we present and describe one possible
accretion--growth mode for planetesimals in perturbed binary star systems,
the so--called Type II runaway growth mechanism.

\section{Planetesimal dynamics in a binary star system: the secular approximation} 
\label{sec:2}

Hereinafter, we will assume that the orbit of the secondary star is
coplanar to the planetesimal disk so that we can neglect 
the inclination of the bodies. This is a reasonable assumption
when the star is close to the primary because of the short 
relaxation time of the disk into the plane of the 
binary orbit. 

Heppenheimer (1978) developed a simplified theory 
for the evolution of the planetesimal eccentricity with time 
based on an expansion of the disturbing function in a power series of the ratio
of the semi-major axes $a/a_b$ where $a$ is the semimajor axis 
of the planetesimal and $a_b$ that of the companion star.
The approximation holds if $a$ is small compared to $a_b$ 
and it stays within the critical semimajor axis $a_c$ for 
dynamical stability derived using direct numerical
integrations by Wiegert and Holman (1997).
By truncating the disturbing function of the secondary star at the second 
order in the eccentricity of the planetesimal, the secular equations 
simplify and lead to the following expressions for the forced eccentricity
$e_F$ and the frequency $u$ of circulation of the perihelion $\varpi$:

\begin{equation}
e_F  =  \frac {5}{4} \, \frac{a}{a_{b}} \, \frac {e_{b}}{1-e_{b}^{2}}
\end{equation}

\begin{equation}
u  =  \frac {3}{4} \pi \frac {1}{(1-e_{b}^{2})^{3/2}} 
m_b \frac {a^{3/2}} {a_b^3} 
\nonumber
\end{equation}

\noindent with

\begin{equation}
\tan(\varpi(t)) = - \frac {\sin (2ut)}{1-\cos (2ut)}\\
\end{equation}

However, Thebault et al. (2006) compared the predictions of the above equations with 
direct numerical simulations and found significant discrepancies, especially
for the oscillation frequency $u$. They empirically 
derived a revised expression for the frequency $u$ which reads:

\begin{equation}
u = \frac {3}{4} \pi \frac {1}{(1-e_{b}^{2})^{3/2}} m_b \frac {a^{3/2}} {a_b^3}
\left [1 + 32 \frac {m_b}{1-e_{b}^{2}}^{3} \left 
(\frac {a} {a_b}^2 \right ) \right ]
\end{equation}
The discrepancy between the numerical and analytical estimate of $u$ is  
reduced from about 70\%, in most of the cases with high $e_b$ and $e$, to less 
than 5\%. 

\begin{figure}
\includegraphics[width=\textwidth]{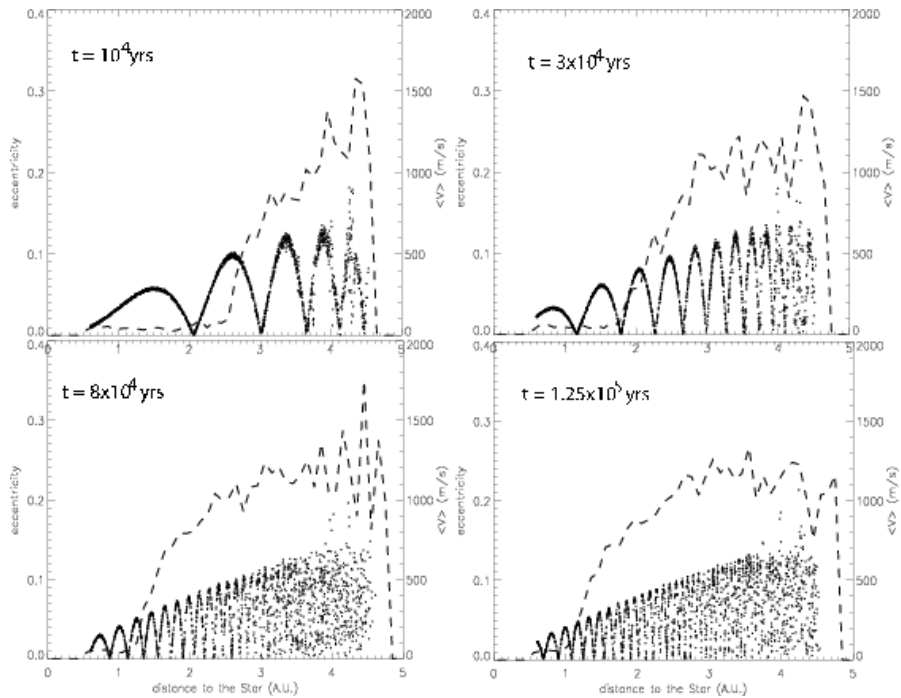}
\caption[]{Evolution of a test particle population perturbed by a stellar
companion having approximately the characteristics of $\gamma$ Cephei:
with $m_{b}=0.25$, $a_{b}=20\,$AU and $e_{b}=0.3$. The dotted line
represents the distribution of average encounter velocities 
between the bodies of the system.
}
\label{numex}
\end{figure}

To illustrate the dynamical evolution of a planetesimal swarm under the
secular perturbations of a companion star we have performed 
a ``pedagogical'' numerical simulation where we integrate the orbits of 
massless test bodies, initially on circular oribts, for about $10^5$ yrs. 
In Fig.\ref{numex} we plot the location of each particle in the
(a,e) plane at different times. The eccentricity of each 
single body oscillates with a period $u$ that depends on the
semimajor axis of the body. If we freeze the dynamical system 
at subsequent times, the slow dephasing of the eccentricity 
of nearby orbits will
translate into a wavy pattern whose density grows with time. 

\noindent
\noindent

\section{Effects of the secular perturbations on $\langle \Delta v \rangle$}
\label{sec:3}

\emph{If} the orientations of the perihelia 
were fully randomized in the planetesimal swarm, the 
effect of the secular perturbations of the companion star on 
$\langle \Delta v \rangle$ would be easy to derive through the
standard relation \citep{lisste93}:
\begin{equation}
\langle \Delta v \rangle \simeq (5/4)^{1/2}  e_F v_{K}
\label{dv0}
\end{equation}
where $v_{K}$ is the local Keplerian velocity.
According to this equation, even very low
values of forced eccentricity would halt accretion. 
In the $\gamma$ Cephei binary system, for example, 
$\langle \Delta v \rangle$ would be approximately 
$1.8 \times 10^3$ m.s$^{-1}$ at 2 AU, the distance from the star where the
planet has been found. This large relative velocity would certainly
prevent mutual accretion in any swarm of kilometer-sized planetesimals
(which have escape velocities of the order of a few m.s$^{-1}$).
In addition, a proper eccentricity component $e_p$ has to be added
to the planetesimal ecentricities,
causing oscillations of the eccentricity around $e_F$ with an
amplitude equal to $e_p$, which leads to even higher impact velocities.
If this was the scenario in the early phases
of planetesimal accumulation in close binary systems,
secular perturbations
pose a serious threat to planetary formation.
However, the predictions of Equ.\ref{dv0} strongly depend on the
assumption that planetesimal trajectories have a randomized 
distribution of orbital angles. This is not true 
at the beginning of the dynamical 
evolution, because secular perturbations 
force of a strong {\it phasing} of the orbits.
The large eccentricity oscillations induced by the companion star
are in phase and associated to a very efficient periastron alignment between
neighboring orbits. This configuration leads to planetesimal encounters 
on almost tangential trajectories
and,  notwithstanding the very high values of forced eccentricity, to low
and accretion friendly {\it relative velocities}, dominated by the Keplerian shear.

At subsequent times, however, the dependence of the precession 
frequency $u$ of $\varpi$ on $a^{1/2} \Delta a$ 
causes a progressive dephasing of nearby orbits.
At some point neighbouring orbits eventually cross
(see the detailed discussion in \citet{the06}), leading
to a sudden and very sharp increase of the encounter velocities.
This is illustrated in
Fig.\ref{numex} where we have superimposed to the (e,a) graph the
corresponding evolution of the average relative velocity
$\langle \Delta v \rangle$. The transition 
between the inner regions, where the orbital phasing
is still strong preventing orbital crossing, and the
outer regions where orbits already cross is very abrupt:
$\langle \Delta v \rangle$ changes from a few m.s$^{-1}$ to about 1km.s$^{-1}$, an
accretion-inhibiting value for kilometer--sized planetesimals, 
within $\simeq 0.1\,$AU. 
\citet{the06} have derived an analytical estimate of
the location in semimajor axis $a_{cross}$ where the orbital 
crossing occurs as a function of time:
\begin{equation}
C_{1}a_{cross}^{1/2} \left[ \left(\frac{a_{cross}}{a_{b}}\right)^{2}+
C_{2}\left(\frac{a_{cross}}{a_{b}}\right)^{4} \right] = \frac{1}{t} ,
\label{realeq}
\end{equation}
where $a_b$, $e_b$ and $m_b$ are the binary semimajor axis, eccentricity 
and mass ratio, respectively, and:
\begin{equation}
C_{1} = \frac {45\pi}{16}  \frac {e_{b}}{\left(1-e_{b}^{2}\right)^{5/2}}
m_{b}  \frac {1}{a_{b}^{2}}
\,\,\, ;
C_{2} = \frac{224}{3} \frac {m_{b}} {\left( 1-e_{b}^{2}\right)^{3} } .
\label{C12}
\end{equation}
Conversely, for a given location $a$ in the planetesimal 
swarm, the time $t_{cr}$ required by the wavefront of high 
impact speeds to reach $a_{cross}$ is:

\begin{equation}
t_{cr} = 6.98 \times 10^2 \frac {(1-e_b^2)^3}{e_b} 
\left ( \frac {m_b}{M_{\odot}} \right )^{1.1} 
\left (\frac {a_b} {10 {\rm AU}} \right )^{4.3} 
\left (\frac {a_{cross}} {1 {\rm AU}} \right )^{-2.8}  \mbox{yr}.
\end{equation}

\begin{figure} 
\includegraphics[angle=0,origin=br,width=1.00\columnwidth]{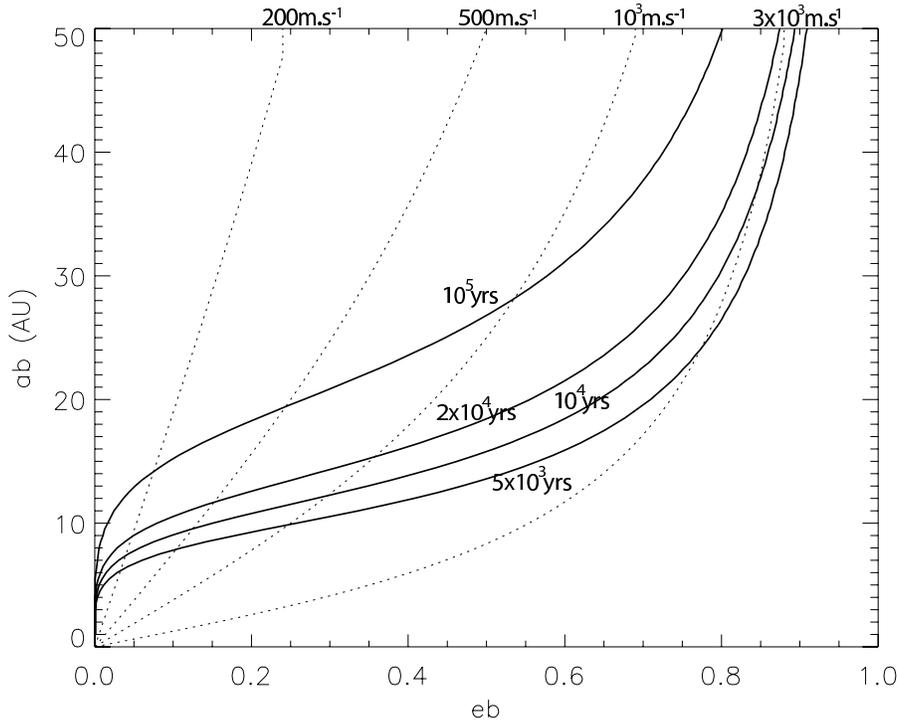}
\caption[]{Value of the minimum companion semi-major axis,
$a_{b.cr1}$, leading to orbital
crossing of planetesimals at 1 AU,
for different values of the crossing time $t_{cr}$,
as a function of the companion 
eccentricity. The companion's mass is fixed, with $m_{b}=0.5$.
The dotted lines represent constant values
of $\langle \Delta v \rangle$ at orbital crossing, as given by Equ.\ref{dv0}
} 
\label{ebab} 
\end{figure}

These analytical formulae hold for $a_b>10\,$AU and for 
$0.05 < e_b < 0.8$. In Fig.\ref{ebab} we show the time required by the
high speed regime to reach 1 AU for for different values of 
$a_b$ and $e_b$ and for $m_b = 0.5$. 
Note that 
for low $e_b$ and large $a_b$, $t_{cr}$ is longer than 
$10^5$ yr, giving a chance to runaway growth to build up large protoplanets 
before an erosion regime is established. 

\noindent
\noindent

An implicit assumption of this scenario is that 
the $initial$ encounter velocities in the planetesimal swarm
are very low, i.e. that the initial orbits
are either circular or fully phased.
This assumption is commonly adopted when addressing
planetesimal accretion in binaries \citep[e.g.][]{hep78,whit98,the04},
but should however be handled with care.
Its validity depends on how and how fast planetesimals decouple from
the gas of the protoplanetary disk to follow their own Keplerian orbits 
around the main star.
If the decoupling occurs almost at the same time for all the
planetesimals and if they are at that moment on almost circular orbits,
then their proper eccentricities $e_p$'s are approximately all equal to $e_F$ and 
the $\varpi$'s are grouped around 0 with consequent low relative
velocities. Basically, it all goes down to how the ``initial'' kilometer--sized
planetesimals are being formed. Unfortunately, so far there are no clues on 
how planetesimals form
in a disk perturbed by the gravity of the secondary star and how
these perturbations may affect their trajectories as they emerge 
from the gas of the disk. 
As a matter of fact, there is not even full agreement on how planetesimals
form around ``normal'' quiet single--stars. This very difficult
exceeds by far the scope of the present, but
there are schematically two competing models for
planetesimal formation in a quiescent protoplanetary disk: 
gravitational instability \citep[e.g.][]{gold73,you02,you04},
where kilometer--sized bodies directly form from small solids in a dense
unstable solid grain layer in the midplane of the protoplanetary disk, 
and collisional-chemical sticking
\citep[e.g.][]{wei80,dom97,dul05}, in which planetesimals are the
outcome of a progressive mutual grain sticking. If gravitational 
instability works out also for disks in binary systems, then
the planetesimal formation timescale is likely to be negligible compared to the
typical timescales for runaway growth and for the onset of fully dephased 
secular perturbations
by the companion star. In
this case, it is reasonable to expect an initial orbital alignment 
within the swarm.  
The decoupling from the gas occurs at a slower rate in
the chemical sticking scenario where planetesimals grow progressively.
They might begin to feel the companion star perturbations 
when they are not yet fully detached from gas streamlines.
However, numerical simulations indicate that the whole growth process 
from grains to kilometer-sized
bodies might not exceed a few $10^{3}$ years \citep{wei00}, i.e. still
shorter than both runaway growth and secular perturbations timescales.

\section{Role of gas drag} 
\label{sec:4}

\begin{figure} 
\includegraphics[angle=0,origin=br,width=1.00\columnwidth]{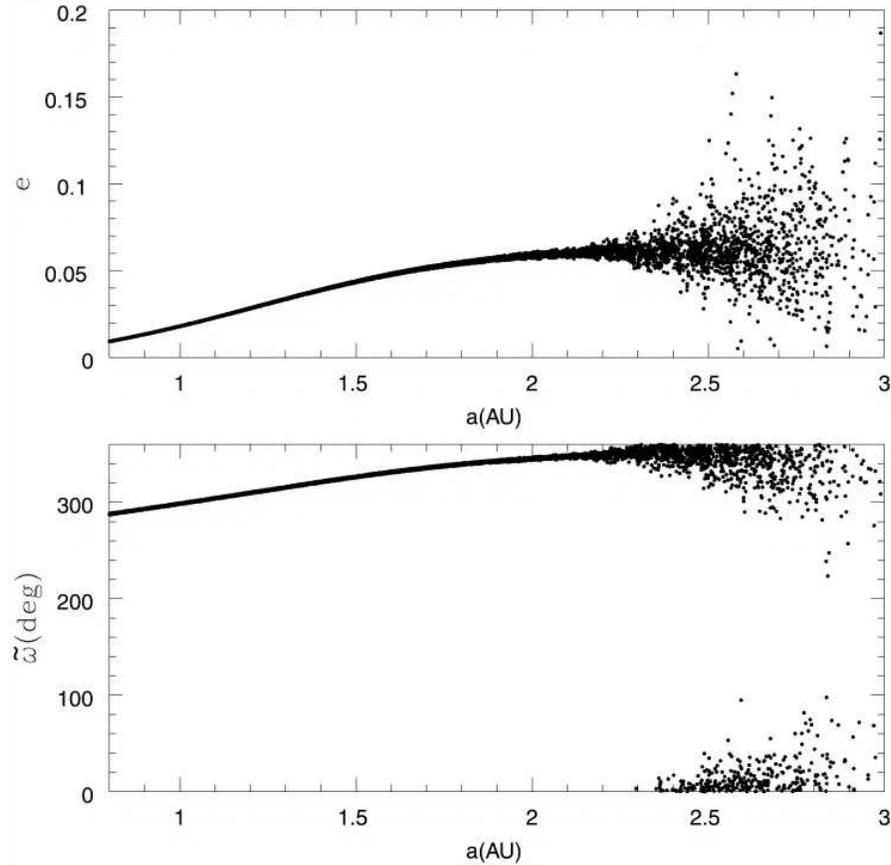}
\caption[]{Periastron and eccentricity distributions of a swarm 
of 5 km diameter planetesimals surrounding the main star of the 
Alpha Centauri system.  After $10^4$ yrs from the beginning of the 
simulation the planetesimal orbits are strongly aligned because of 
the coupling between gas drag and secular perturbations of the
companion star.
} 
\label{gasex}
\end{figure}

So far, all effects other than the gravitational perturbations
of the companion star have been neglected.
However, in the standard
planetary formation scenario, the initial stages of
planetesimal accumulation  are believed to take place in presence
of significant amounts of primordial gas. 
Marzari and Scholl (2000) have shown that 
gas drag damps planetesimal eccentricities and also
restores a strong perihelion alignment for equal size 
bodies. If the density of the gas is large enough 
all the $\varpi$s of planetesimals halt their circulation and 
tends towards a fixed value $\varpi_{al}$ equal to $270^{\circ}$.
This tends to cancel out the large eccentricity oscillations
forced by the companion.
In Fig.\ref{gasex} we show this phenomenon for 5 km planetesimals embedded
in a protostellar disk similar to the minimum mass solar nebula and
perturbed by an outer star. The parameters of the binary 
system recall
those of $\alpha$ Centauri. In this simulation and in all the 
following ones, the drag force is modelled in laminar gas 
approximation as

\begin{equation}
\ddot{\bf r}  = - K v_{rel} {\bf v_{rel}},
\end{equation}
where $K$ is the drag parameter given by $K=3 \frac{\rho_g C_d} 
{8 \rho_{pl} s}$ \citep{karo} where $s$ is the radius of the planetesimal,
$\rho_pl$ its mass density, $\rho_g$ the gas density of the 
protoplanetary disk and $C_d$ a dimensionless drag coefficient
related to the shape of the body ($\sim 0.4$ for spherical bodies)
\footnote{Our gas drag model is a simplified one where the gas disk
is assumed to be fully axisymmetric and follows a classical
\citet{haya81} power law distribution. It is however more
than likely that in reality the gas disk should depart
from this simplified view because it would also "feel" the
companion star's perturbations. Several numerical studies
have investigated the complex behaviour of gaseous disks in binary
systems. They all show that pronounced spiral structures
rapidly form within the disk \citep[e.g.][]{arty94,savo94}
and that gas streamlines exhibit radial velocities.
To follow the dynamical behaviour of planetesimals in such non-axisymmetric
gas profiles would require a study of the $coupled$ evolution
of both gas and planetesimal populations, have to rely on hydro--code modeling
of the gas in addition to N--body type models for the planetesimals.
Such an all--encompassing gas$+$planetesimals modeling is clearly the next step
in binary disk studies and have already been started by several teams.
It is interesting to note that preliminary results seem to show that
planetesimal behaviours in systems with ``realisitic'' gas disk modeling
do not seem to drastically depart from the behaviour in the axisymmetric case.
There is in particular no phase alignment between eccentric planetesimal orbits
and gas streamlines, so that gas friction on planetesimals is still
very high (Paardekooper, private communication).}

The orbits of the planetesimals 
are all aligned for an indefinite time up to about 2 AU from 
the star and the $\langle \Delta v \rangle$ is
of the order of a few meters per second. Under this condition, 
planetesimal collisions lead to accretion all the time. 

\noindent
\noindent

\begin{figure} 
\includegraphics[angle=0,origin=br,width=1.00\columnwidth]{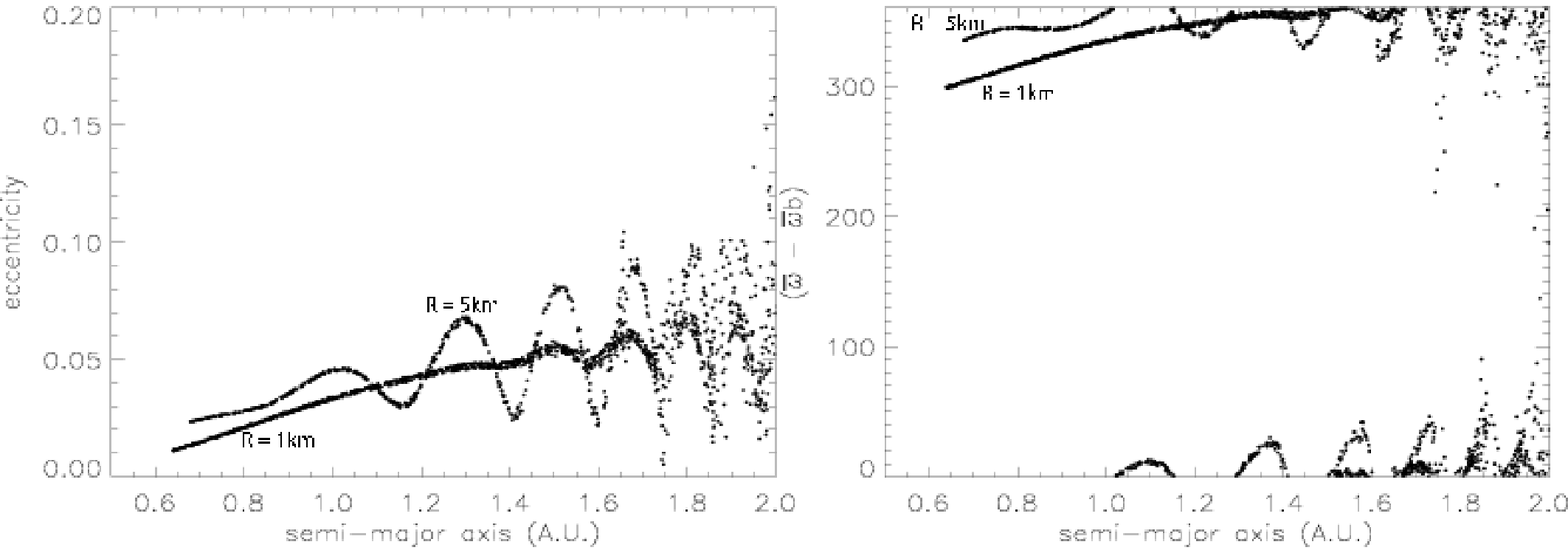}
\caption[]{Example gas drag run.
Snapshots, at $t=3\times10^{3}$yrs, of the ($e,a$) and ($\varpi-\varpi_b,a$)
distributions for 2 planetesimal populations of size $R_1=1\,$km and $R_2=5\,$km.
$\varpi-\varpi_b$ is the difference, in angular degrees,
between the particles and the companion star's longitude of periastron.
The companion star orbital parameters are: $a_b = 10\,$AU, $e_b = 0.3$, $m_b = 0.5$
 } 
\label{exsmall} 
\end{figure}

Unfortunately, this alignment depends on the balancing between 
the drag force and the strength of the secular perturbations. 
For a lower gas density or for a larger size of the planetesimals
the perihelia moves away from $\varpi_{al}$ towards larger 
values and the alignment is perturbed by a wavy pattern that 
forewarns the onset of a regime dominated by secular perturbations. 
This behaviour is illustrated in  Fig.\ref{exsmall} for a bimodal planetesimal
population characterized by two different sizes. 
\noindent
\noindent

The different alignment of planetesimals according to their 
size has a crucial impact on the accretion process. 
While equal size planetesimals keep a low $\langle \Delta v \rangle$ since their
orbits are oriented towards the same angle, 
when different size planetesimals encounter each other 
the relative velocity is large most of the time because of 
the forced unpaired
orbital alignment. The high eccentricities induced by the 
secular perturbations introduce a large non-tangential 
component in the relative velocity that easily overcomes
the Keplerian shear and cause high velocity impacts. 
In Tables I and II we report the average encounter velocities 
$<\Delta v(R_1,R_2)>$
between planetesimals of size $R_1$ and $R_2$ 
for two different binary star orbital configurations.
The assumed density of the gas drag corresponds to that of 
the Minimum Mass Solar Nebulae derived by \citet{haya81}.
The velocity values have been 
obtained with numerical integrations of the planetesimal trajectories
over $2\times 10^{4}$years. 
For the case displayed in Table I, it is worth noticing that,
in absence of the gaseous component, 
the parameters of the binary system would lead to orbital 
crossing of nearby planetesimal orbits in less than $\simeq 5\times 10^{3}$years 
(see Fig.\ref{ebab}). The swarm would stop growing and erosion 
would take place.  When we include the effects of gas drag 
the $\langle \Delta v \rangle$ remains always low for 
equal size planetesimals. However, for bodies of different sizes, 
the different phasing caused by the coupling between gas
drag and secular perturbations pump up the relative velocities. 

\noindent
\begin{table} 
\caption[]{Average encounter velocities in m.s$^{-1}$, at 1\,AU from the primary,
within a population of ``small'' planetesimals $1<R<10\,$km for a
gas drag simulation with the companion star parameters
$m_b=0.5$, $a_b=10\,$AU and $e_b=0.3$
$m_b=0.5$, $a_b=20\,$AU and $e_b=0.4$.
$\Delta v_{R1,R2}$ values are
averaged over the time interval $0<t<2\times10^{4}$yrs.
Initial starting encounter velocities are such as $\Delta v_0\simeq 10$m.s.$^{-1}$.
$\Delta v$ values in bold correspond to accreting impacts for
all tested collision outcome prescriptions. Underlined values
are those for which we obtain different accretion vs. erosion
balance depending on the tested prescription. Values in classical roman
characters correspond to cases for which all tested models
agree on a net erosive outcome.
}
\label{dvmatrix1} 
\begin{tabular}{c | cccccccccc}\\
\hline 
Sizes (km) & 1 & 2 & 3 & 4 & 5 & 6 & 7 & 8 & 9 & 10\\ 
\hline 
1 & {\bf10} & 154 & 233 & 285 & 327 & 360 & 391 & 426 & 452 & 458\\ 
2 & 172 & {\bf10} & 94  & 133 & 187 & 223 & 262 & 287 & 316 & 334\\ 
3 & 238 & 84 & {\bf11}  & 54  & 99 & 137 & 177 & 200 & 230 & 254\\ 
4 & 289 & 144 & 63  & {\bf12} &\underline{40}& 80  & 115 & 149 & 171 & 198\\ 
5 & 325 & 188 & 103 &\underline{43}& {\bf12} & \underline{32}& \underline{70}& 100 & 122 & 154\\ 
6 & 373 & 228 & 144 & 83  & \underline{32}& {\bf11} & {\bf36}& \underline{56}& 84  & 104\\ 
7 & 400 & 261 & 182 & 113 & \underline{68}& {\bf36}& {\bf12} & {\bf35}  &\underline{48} & \underline{76}\\ 
8 & 428 & 298 & 212 & 147 & 98 & \underline{56}&{\bf35}& {\bf12} &{\bf36}& \underline{45}\\ 
9 & 450 & 310 & 238 & 168 & 123 & 83  & \underline{48}& {\bf36}& {\bf13} & {\bf31}\\ 
10& 453 & 338 & 263 & 196 & 152 & 107 & \underline{73}& \underline{48}& {\bf31} & {\bf13}\\ 
\end{tabular} 
\end{table}
\noindent
\begin{table} 
\caption[]{Same as Table \ref{dvmatrix1}, but with companion
star parameters $m_b=0.5$, $a_b=20\,$AU and $e_b=0.4$.
}
\label{dvmatrix2} 
\begin{tabular}{c | cccccccccc}\\
\hline 
Sizes (km) & 1 & 2 & 3 & 4 & 5 & 6 & 7 & 8 & 9 & 10\\ 
\hline 
1 & {\bf11} & 127 & 204 & 255 & 298 & 342 & 368 & 390 & 417 & 442\\ 
2 & 126 & {\bf10} & 84  & 139 & 185 & 227 & 258 & 290 & 317 & 340\\ 
3 & 200 & 91 & {\bf11}  & 68  & 108 & 158 & 186 & 218 & 246 & 272\\ 
4 & 258 & 146 & 61  & {\bf9} &\underline{48}& 88  & 120 & 154 & 186 & 209\\ 
5 & 301 & 192 & 113 & 54  & {\bf12} & \underline{44}& \underline{75}& 111 & 136 & 164\\ 
6 & 339 & 232 & 152 & 99  & \underline{43}& {\bf10} & {\bf31}& \underline{66}& 92  & 119\\ 
7 & 361 & 262 & 181 & 126 & \underline{77}& {\bf28}& {\bf13} & {\bf26}  &\underline{56} & 87\\ 
8 & 395 & 295 & 219 & 159 & 112 & \underline{68}&{\bf26}& {\bf11} &{\bf28}& \underline{48}\\ 
9 & 425 & 320 & 246 & 190 & 136 & 92  & \underline{55}& {\bf25}& {\bf11} & {\bf23}\\ 
10& 446 & 346 & 266 & 208 & 163 & 122 & 82  & \underline{49}& {\bf22} & {\bf12}\\ 
\end{tabular} 
\end{table}

\section{Dependence of the accretion process on the binary parameters.}
\label{sec:5}

Once we determine the $\langle \Delta v_{(R_1,R_2)} \rangle$ for any pair of 
planetesimal size we must compare it with the threshold velocity
$v_{ero}$ for that pair of projectile and target 
to test whether either erosion or 
accretion will be the result of the collision. 
As pointed out in \citet{the06}, in literature one can find 
many prescriptions of the collisional outcome between two 
planetesimals given their impact velocity. Adopting the 
approach outlined in \citet{the06}, three different values of 
$v_{ero}$ are computed according to \cite{hol94}, \cite{mar95} and
\cite{benz99}. Erosion (or accumulation) is assumed to occur when 
$\langle \Delta v_{(R_1,R_2)} \rangle$ is higher (or lower)
than all the three values 
of $v_{ero}$. In the intermediate cases we prefer not to draw any
definitive conclusion letting them as undefined cases. 
In Tabs.\ref{dvmatrix1}\&\ref{dvmatrix2} 
a different text font is used to indicate when the
computed  $\langle \Delta v_{(R_1,R_2)} \rangle$
leads to accretion, erosion, or to an undefined situation. 
By inspecting the tables, it is interesting to note that anytime we depart
from the diagonal stripe $R_1=R_2$ there is a sharp increase of 
$\langle \Delta v \rangle$, almost always exceeding 
the threshold value for accretion, at least for the binary
parameters adopted in the tables. 

The knowledge of the relative velocities between different size planetesimals
is the first step towards fixing constraints on the dynamical environment
allowing planetary formation. These velocities should be used in 
the context of a full numerical model for planetesimal accretion
that takes into account also the frequency with which different or
equal size planetesimals collide. 
If the size distribution 
of the swarm privileges impacts between equal size planetesimals, 
it comes out clearly from the data in the tables that accretion is 
possible also for very close and eccentric binary systems. 
On the other hand, if collisions between different size planetesimals
dominate the evolution of the planetesimal swarm, then the 
different alignment of perihelia might halt the accretion process. 
Unfortunately, the size distribution of a planetesimal swarm
is poorly constrained at the beginning of the accretion process 
and it may also be a misleading concept \citep{wetina00}.
The frequently adopted assumption that ``initial'' planetesimals
have a given size $R_{pl}$ is numerically convenient but certainly
an oversimplification. Regardless of the details of
the planetesimal formation process, there have to be a dispersion $\Delta R_{pl}$ in
planetesimal sizes. The amplitude of $\Delta R_{pl}$ is difficult
to estimate, but it seems likely that it spans over at least one 
(if not several) order(s) of magnitude \citep[see for instance Fig.3 of][]{wei00}.
If this is the case, then in the binary stars cases explored here,
encounters leading to erosion (``NA'' type encounters)  might
largely dominate over those leading to accretion (type ``A''). We have
checked this by performing
a simplified test where we assume for the planetesimal size distribution 
a Gaussian function centered on $R=5\,$km of variance $\Delta R_{pl}^2$.
By counting the number of A--type encounters vs. NA--types using the 
values of $\langle \Delta v \rangle_{(R_1,R_2)}$ reported in 
Tabs.\ref{dvmatrix1} and \ref{dvmatrix2} we find that accretion dominates
over erosion only
when $\Delta R_{pl}< 0.8\,$km. This value of $\Delta R_{pl}$ 
appears to be small, possibly too small for 
any realistic initial size distribution.
As a consequence, gas drag tends to prevent, or at least
slow down, accretion in the initial stages of planetesimal 
evolution.  
As accretion 
proceeds, larger bodies are formed and impacts occur more and more 
frequently between different size bodies.  The different phasing 
caused by gas drag for different size planetesimals increases 
the number of NA--type encounters and erosion dominates, 
at least for the binary parameters
given in the tables. 

\begin{figure} 
\includegraphics[angle=0,origin=br,width=1.00\columnwidth]{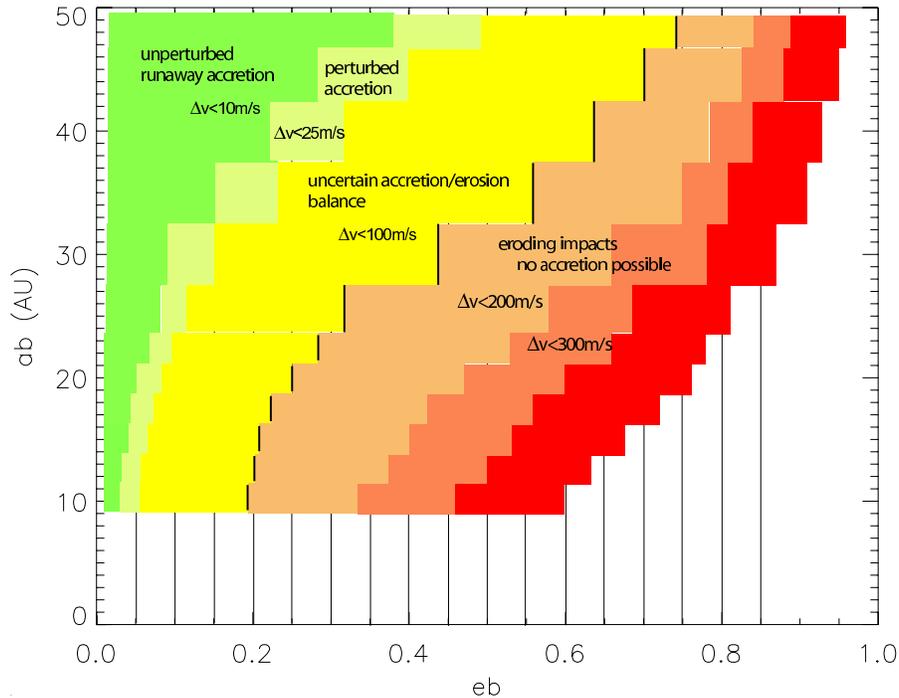}
\caption[]{Encounter velocities averaged, over the time interval
$0<t<2\times10^{4}$yrs, between $R_1=2.5$ and $R_2=5\,$km bodies
at 1 AU from the primary star, for different values of the
companion star semi-major axis and eccentricity.
The short black vertical segments mark the limit beyond which
$<\Delta v_{(R1,R2)}>$ values correspond to eroding impacts
for all tested collision outcome prescriptions.
 } 
\label{dv5-2.5} 
\end{figure}

\begin{figure} 
\includegraphics[angle=0,origin=br,width=1.00\columnwidth]{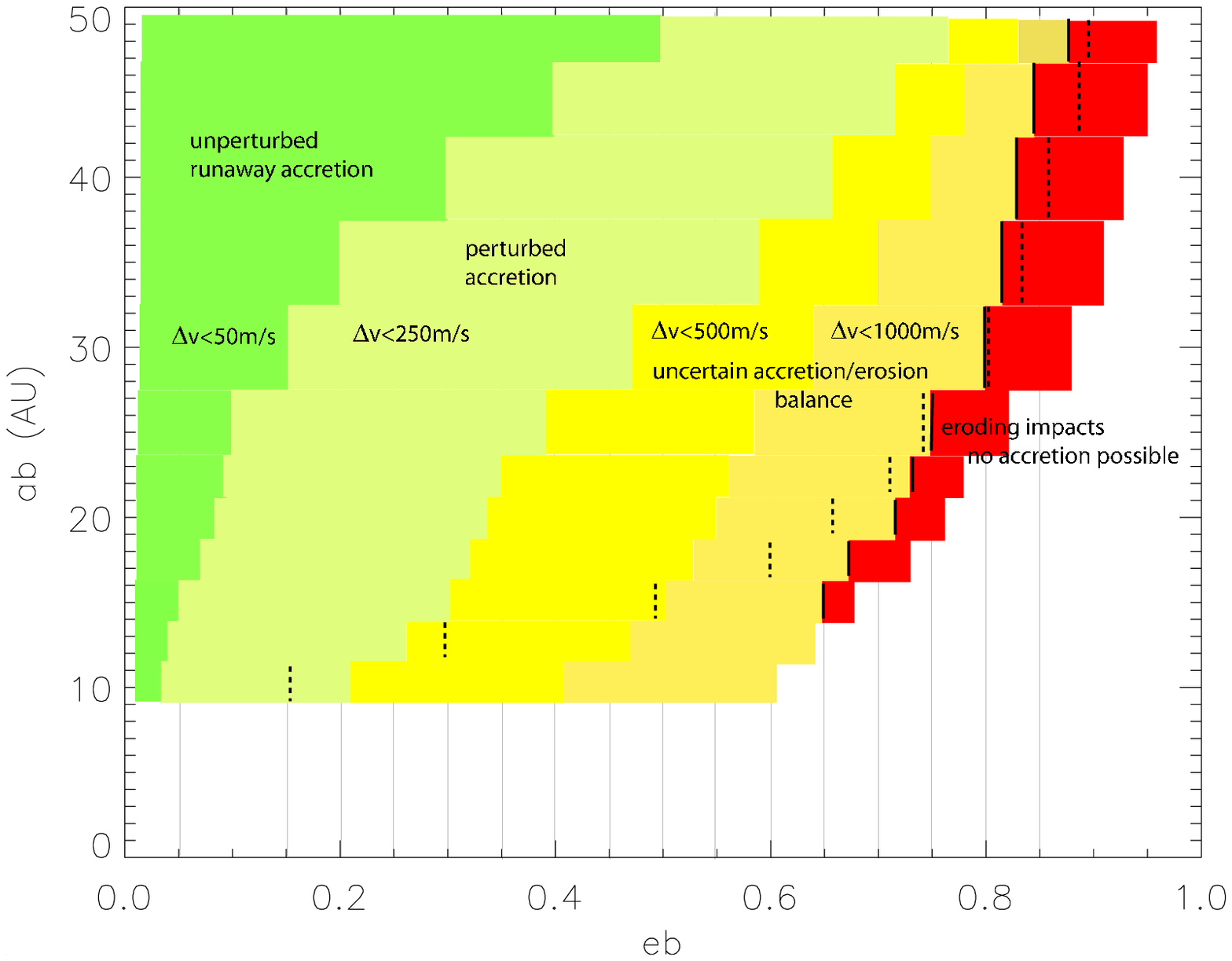}
\caption[]{Same as Fig.\ref{dv5-2.5}, but for $R_1=15$ and $R_2=50\,$km bodies.
The short black vertical segments mark the limit beyond which
$<\Delta v_{(R1,R2)}>$ values correspond to eroding impacts
for all tested collision outcome prescriptions.
The short dashed vertical segments mark the position beyond which orbital crossing
occurs, despite the effects of gas drag, for 50\,km bodies.
 } 
\label{dv50-15} 
\end{figure}

The limit of a purely dynamical approach that estimates 
the planetesimal relative velocity but neglects their size distribution
can be partly overcome by taking into
account that on average the projectiles delivering the maximum 
kinetic energy to a target of size $R_2$ are roughly those of size 
$R_1 \simeq 1/2 R_2$ \citep[see][]{the06}.
With this assumption we can probe the feasibility 
of accretion by testing $\langle \Delta v \rangle$ against $v_{ero}$ 
only for a limited number of different size body pairs. 
By exploiting this simplification,  \citet{the06} have 
tested the chances of planetesimal accretion for 
120 
different binary systems with semimajor axis $a_B$ ranging from 
10 to 50 AU and eccentricity from 0.05 to 0.9. The mass ratio
has been kept fixed to $m_b=0.5$. In all the simulations
only two size pairs have been 
considered,  $R_1=2.5$km vs. $R_2=5$km, representative of small
planetesimals, and $R_1=15$km vs. $R_2=50$km, representative 
of intermediate size bodies in the swarm. 
In Figs.\ref{dv5-2.5}and\ref{dv50-15} different color coding are
used to outline the planetesimal collisional fate, depending on the 
binary parameters:

\begin{itemize}
\item
1) {Dark green}. The evolution of a planetesimal disk 
in binaries belonging to these regions
is not significantly affected by the companion 
star perturbations. $\langle \Delta v \rangle$ remains low and 
accretion should proceed as in the standard model for single stars. 
\item
2) {Light green}. Within these regions, $\langle \Delta v \rangle$ is 
increased by the gas drag coupling to the secular perturbations 
but remains below $v_{ero}$. Accretion is still possible but
the amplitude of the gravitational focusing factor 
$(v_{esc(R_2)}/\Delta v)^{2}$ is significantly reduced. 
This slows down the
accretion rate compared to the single--star unperturbed case.
\item
3) {Yellow}. Here $\langle \Delta v \rangle$ is comparable
to the erosion velocities computed with the different 
prescriptions adopted at the beginning of the study for 
$v_{ero}$. This is a limiting situation where the preference 
of the system towards accretion or erosion depends on the 
details of the still not well defined collisional physics. 
We prefer not to draw 
any definite conclusion in this case. 
\item
4) {Orange and Red}. In these regions, $\langle \Delta v \rangle$
exceeds $v_{ero}$ for all three different collision outcome
prescriptions considered. Collisions always result in a net mass loss
and mutual accretion is impossible.
\end{itemize}

As expected, systems with higher eccentricity or lower binary separation 
are more critical for planetesimal accretion. An empirical fit has been 
performed in \citet{the06} to outline the region in the 
($a_b$,$e_b$) phase space where accretion is possible for the two different 
size ranges of planetesimals:  
\begin{equation}
e_{b} \simeq 0.013 \left(\frac{a_b}{10{\rm AU}}\right)^2
\,\,\,{\rm,\,\,for\,\,small\,\,\,1<R<10km\,\,\,planetesimals,}
\label{ebab1}
\end{equation}
and
\begin{equation}
e_b \simeq 0.018 \left(\frac{a_b}{10{\rm AU}}\right)^2
\,\,{\rm,\,\,for\,\,large\,\,\,10<R<50km\,\,\,planetesimals.}
\label{ebab2}
\end{equation}
By extrapolating the fit to large values of  $a_b$ one can figure out 
that for binary separations $a_b \geq 90$AU the planetesimal 
accretion process is not significantly perturbed by the companion
star gravity. 

One interesting feature is the extent of the
``light--green'' region where planetesimal 
accretion is not halted but possibly slowed down since the 
gravitational focusing factor is reduced. As a consequence,
``classical'' runaway growth is not possible and might be replaced
by the alternative, possibly slower type II runaway mode identified by
\citet{kort01}.


\section{Type II runaway growth}
\label{sec:6}

In this section we describe details of planetesimal accretion simulations 
in the Sun-Jupiter ``binary'' system \citep{kortenkamp2000,kort01}.  
This system has traditionally been used for
initial characterizations of dynamical effects in binary-star systems
\cite[e.g.][]{hep78,whit98}.  In the models described here we use units
of binary separation $D$ rather than AU.  Note also that the modeling 
included Saturn as well (a triple system) but we omit Saturn from most of the discussion, referring
simply instead to the Sun-Jupiter ``binary.''

Initial orbits for the
planetesimals were circular, coplanar with the nebular midplane, and
uniformly distributed from 10-50\% of $D$.  As the system evolves the
combination of gas drag and secular perturbations from the companion
Jupiter (and Saturn) leads to a size-dependence in $e$ and $I$ as well
as in the phasing of the orbital orientation angles, as described earlier 
in this review chapter.  Note that the 
system is fully three dimensional.  The companion orbits are inclined 
with respect to each other and the protoplanetary disk.

As described above, the distribution of encounter velocities is a critical factor which
determines how a population of planetesimals will accumulate into
larger bodies.  High encounter velocities can actually lead to impact
erosion of the target rather than growth.  On the other hand, low
encounter velocities can gravitationally enhance the collision
cross-section of the target \citep{opik1951}. The effective
cross-section becomes
\begin{equation} \label{eq:grav_focus}
\sigma_{\msc{eff}} = \sigma \left( 
                        1 + \frac{v_{\msc{esc}}^2}{v_{\msc{enc}}^2} 
                            \right) \ ,
\end{equation}
where $\sigma$ is the target's geometric cross-section,
$v_{\msc{enc}}$ is the projectile's encounter velocity, and
$v_{\msc{esc}}$ is the combined surface escape velocity of the target
and projectile.  

The size-dependent phasing of orbital elements leads
to low encounter velocities between similar size bodies and high
encounter velocities between bodies of markedly different size.  For
the Sun-Jupiter model described above ''binary'' encounter velocities were
calculated for all intersecting orbits over the
entire range of planetesimal sizes.  \,Figure~\ref{typeII_runaway_fig1}
shows examples of these en- 

\noindent
\begin{minipage}{10cm} 
        \refstepcounter{figure}
        \label{typeII_runaway_fig1}
        \sloppy  
        \vspace{0cm} 
        \centerline{  
          \hspace{-4.75cm}
          \scalebox{0.5}{\includegraphics{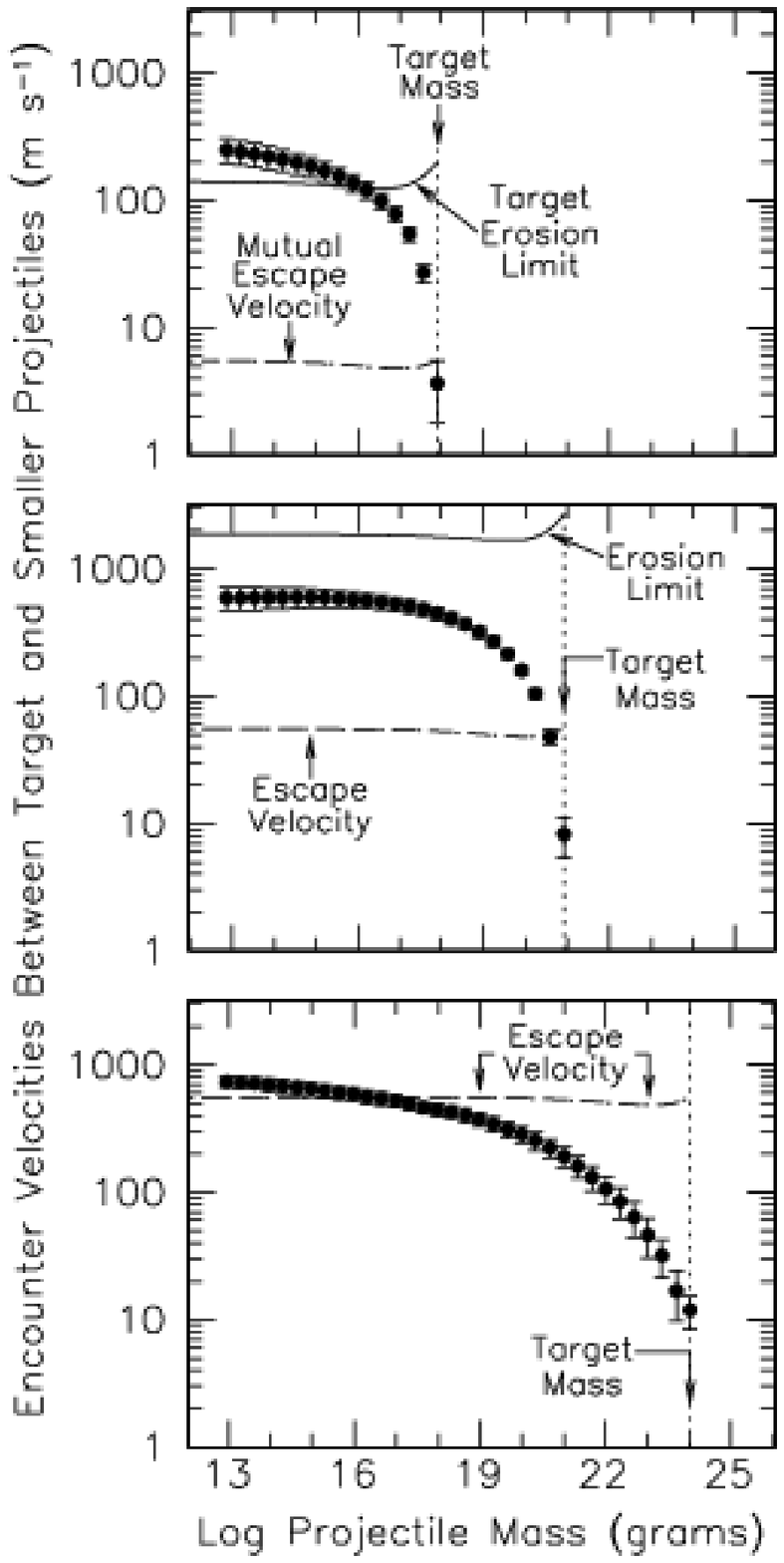}}}
        \hspace{4.75cm}
        \parbox{7cm} {
          \small
          \makebox[0cm]{}\\[-10cm]
          {\bf Fig.~\ref{typeII_runaway_fig1}.} Mean encounter velocities ($\pm \sigma$) 
are shown for various size planetesimals with intersecting orbits near 33\% \textsl{D}, 
where \textsl{D} is the Sun-Jupiter separation distance.  
The size-dependent orbital phasing of bodies on intersecting orbits leads to high 
encounter velocities between bodies of markedly different sizes.  The surface escape 
velocity of the combined target/projectile is shown by the horizontal dashed line.  
In the top two panels the erosion limit is indicated by the horisontal solid line.  
For encounter velocities above the erosion limit impacting projectiles crush and eject 
more than their own mass (crushing strength was assumed to be $10^8$ ergs/g).  
Top: Encounter velocities for $10^{18}$ g targets and projectiles of equal 
or lessor mass. Middle: Targets with masses of $10^{21}$ g are large enough 
that impacts of all projectiles lead to growth.  Bottom: Planetsimals capable of growing to $10^{24}$ !
 g have surface escape velocities near 500 m\,s$^{-1}$.  Encounter velocities between these large planetesimals and bodies as much as 8-10 orders of magnitude less massive are below the escape velocity, allowing for significant gravitational focusing.}
\end{minipage}

\noindent
counter velocities for various target bodies
and equal or lesser size projectile bodies with orbits near 33\% $D$,
where, again, $D$ is the Sun-Jupiter separation distance.

\begin{figure}[!ht] 
\vspace{-2cm}
\includegraphics[angle=0,origin=br,width=1.00\columnwidth]{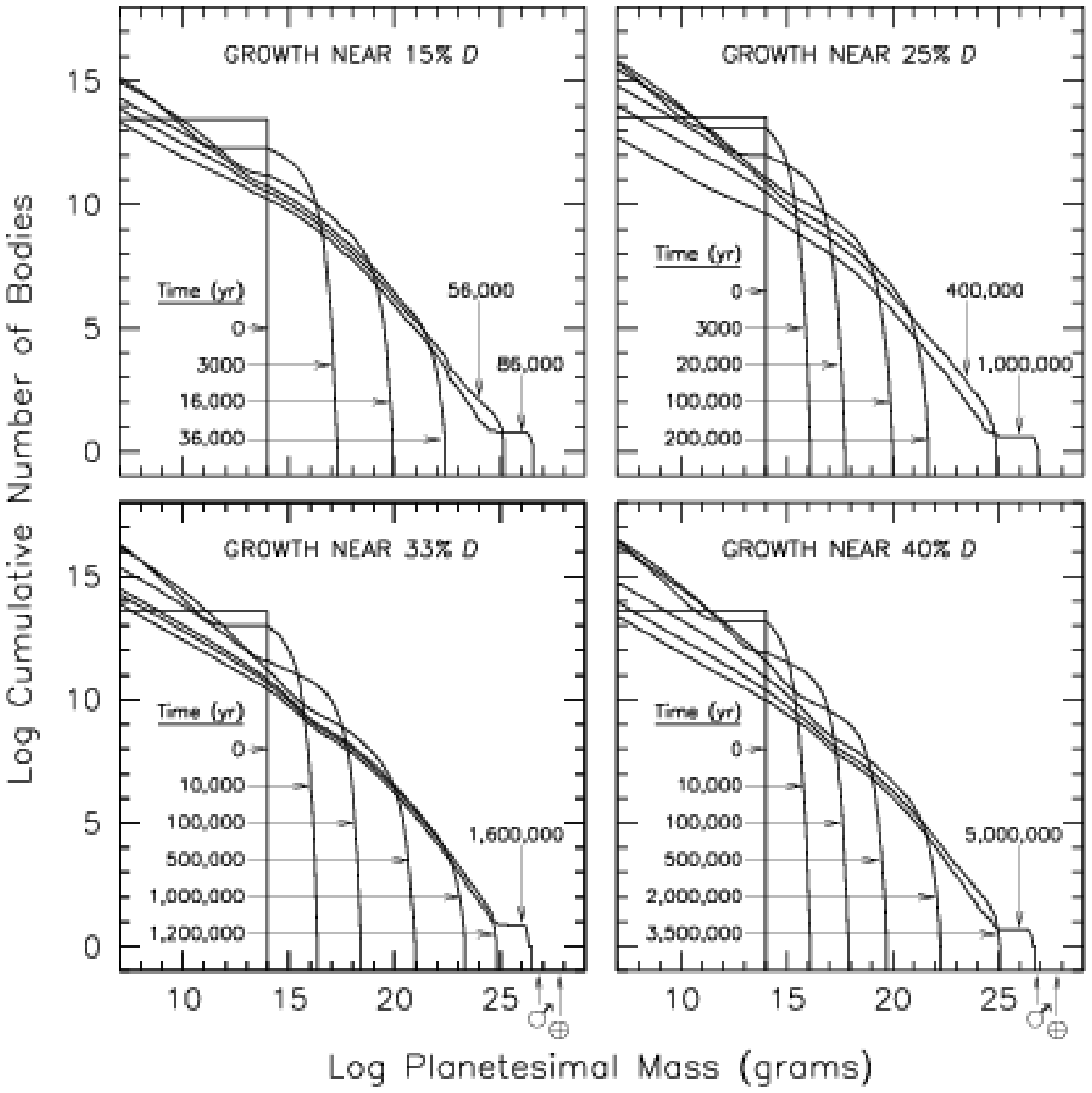}
\vspace{-1.5cm}
\caption[]{Growth of
        planetesimals in a circumsolar disk perturbed by the
        ``companion'' Jupiter at a distance \textsl{D} from the sun.  The
        habitable zone, simply defined here as the Venus-Mars region,
        stretches from about 10-25\% of \textsl{D}.  Initially all planetesimals have identical
        masses of 10$^{14}$ grams.  The initial surface density
        of planetesimals at 15\% of \textsl{D} is 10 g cm$^{-2}$, scaling
        as $r^{-3/2}$ with heliocentric distance $r$.  This is
        roughly consistent with the so-called ``minimum mass'' surface
        density.  Collision fragments smaller than 10$^{7}$ g
        ($\sim$1 meter) are presumed lost via nebular gas drag and
        therefore removed from the simulation.  Growth is calculated
        in four different regions centered on 15, 25, 33, and 40\% of
        \textsl{D}.  For comparison, the masses of Earth and Mars are
        indicated by their respective symbols.  The top two panels represent growth in the
        habitable zone and show initial orderly growth transitioning
        to Type II runaway growth of Mars-size planetary embryos in
        $10^5$ to $10^6$ years.  Beyond the habitable zone (bottom
        panels) growth is slower but eventually also produces Mars-size embryos.  Figure is adapted from \citet{kort01}.
 } 
\label{typeII_runaway_fig2} 
\end{figure}

Modeling the growth of planetesimals in this scenario is considerably
more difficult than tracking their orbital evolution.  Full-scale
$N$-body simulations of planetesimal growth that include mutual
gravitational perturbations, secular perturbations, and gas drag are
beyond the reach of current techniques.  Theoretically one would need
to include $\sim$10$^{12}$ small ($\sim$10$^{14}$ g) planetesimals to
form a single $10^{26}$ g terrestrial planet embryo.  Direct $N$-body
integrations of mutually perturbing planetesimals cannot even remotely
approach this figure, treating only about $\sim$10$^4$ bodies over the time
scale required.  On the other hand, existing statistical simulations
of planetary growth are not limited by the number of bodies.  However,
these simulations assume that the orbits are completely randomized so
they cannot easily accommodate the size-dependent orbital evolution
described above.   To overcome these shortcomings, \citet{kort01} developed a hybrid approach that capitalizes on the strengths of
each technique.  They used $N$-body integration to
map the size-dependent velocity distributions.  These velocity
distributions are then used in modified statistical accretion
simulations to model the collisional accumulation of the
planetesimals.  This approach led to the identification of an
alternative form of runaway growth that is facilitated by the secular
perturbations of massive companions \citep{kort01}.  
Figure~\ref{typeII_runaway_fig2} shows results from these growth simulations.

Four separate simulations were performed, centered on 15\%, 25\%, 33\%
and 40\% of $D$.  For comparison, the habitable zone extends from
about 10-25\% of $D$ (crudely defined as the Venus-Mars
region).  In all four regions the growth is characterized as
``orderly'' \cite[non-runaway, see][]{saf69} until the distribution
reaches approximately 10$^{24}$ g, or about  the size of the largest
asteroid---(1)~Ceres.  The planetesimal size distribution then becomes
bi-modal, transitioning to runaway growth and producing Mars-size
embryos.  All planetesimals were assumed to be uniformly distributed
across a region and any bodies separated by more than 10 mutual Hill
radii were considered dynamically isolated \citep{chambers1996} and
not allowed to collide.  This resulted in multiple runaway embryos
emerging in each region.

Note that mutual perturbations between planetesimals were not included
in these calculations.  This new form of runaway growth arises when
secular perturbations and gas drag act together to establish
size-dependent encounter velocities that remain low when colliding
bodies are of similar size.  Collisions between bodies of
significantly different size are at high velocity and can lead to
cratering and erosion, but these simulations show that growth overcomes
erosion (Fig.~\ref{typeII_runaway_fig2}).  This alternative mode of growth, 
which was eloquently dubbed
``Type II'' runaway growth, allowed for the formation of terrestrial
planet embryos throughout the habitable zone of the Sun-Jupiter ``binary.''  
This general result should apply
regardless of whether the perturbations are 
from Jupiter-like
companions formed earlier by disk-instability \cite[e.g.][]{bos1,mayer2002}, 
stellar-mass objects in
multiple-star systems, or perhaps even short-lived instabilities that lead to
asymmetries in protoplanetary disks.

Classical ``Type~I'' runaway growth occurs in a self-gravitating
population of planetesimals.  Random orbital kinetic energy is
exchanged during gravitational encounters between large and small
bodies and the population trends towards energy equipartition, a
process dubbed ``dynamical friction'' \citep{stewart1980}.  Dynamical
friction  lowers the encounter velocities of the larger bodies with
respect to each other, enhancing their effective collision
cross-sections and increasing the rate at which they accumulate each
other.  Under these conditions nearly the entire growth period up to
embryo-size is in Type~I runaway mode.  In our simulations, which are
not self-gravitating, the size-dependent phasing of orbital elements
holds encounter velocities low between all similar-size bodies
(typically 1 to 10 m\,s$^{-1}$, see Fig.~\ref{typeII_runaway_fig1}).  Initially these
encounter velocities exceed the planetesimal escape velocities so
there is no enhancement of collision cross-sections and growth is
orderly.  As larger and larger bodies grow their escape velocities
approach and then exceed the relatively low encounter velocities,
causing the transition from orderly growth to Type~II runaway growth.
In this way, the effects of dynamical friction are mimicked by the
size-dependent phasing of orbital elements.

\section{Conclusions}
\label{sec:7}

The process of planetary formation around the primary 
star of a binary system is complicated
in all its stages by the gravitational perturbations 
of the companion star. However, the existence of 
some gas giant planets in binary systems with separation 
of a few tens of AU suggests that these perturbations 
are not always strong enough to prevent the formation of 
a planet. We have analysed within the standard model 
of planet growth how the secular perturbations 
of the companion star affect the various stages of
planetesimal accretion. Ours studies show that 
when we include also the effects of gas drag on the 
motion of small planetesimals, their orbits become 
phased both in eccentricity and perihelion longitude
\citep{mascho00}.
At a first sight, this alignment favors fast accretion 
keeping low the encounter velocities between the 
bodies in spite of the large values of eccentricity forced
by the secular perturbations.
However, the angle towards which the perihelia align depends 
on the size of the planetesimals and on their 
distance from the star \citep{the04,the06}.
In the earlier stages of planetesimal accretion all the 
bodies have roughly the same size and the encounter velocities,
because of the favorable phasing, are low. 
However, as soon as larger bodies emerge in the population
the different alignments of the orbits lead to large impact 
velocities that may prevent the onset of runaway growth or even
cause erosion of the bodies inhibiting the formation of a planet. 
Through a full numerical approach we have analysed to which 
extent the different alignment influences the random planetesimal
velocities. We have also mapped the values of the binary orbital 
parameters (semi-major axis and eccenticity)
for which accumulation dominates over
erosion despite of the secular perturbations, or viceversa. 
Our modelling shows that for binary of separations $a_b\leq40$AU, only
very low $e_b$ allow planetesimal accretion to proceed as
in the standard single-star case. On the contrary, only relatively
high $e_b$ values (of at least 0.2 in the closest $a_b=10$AU sepration
explored and at least $\sim 0.7$ for $a_b=40$AU) lead to
a complete stop of planetesimal accretion.
In most cases, when the perturbations of the massive companion
on the planetesimal disk is significant but not strong enough
to halt accretion, runaway growth can still occur, but in a different
way respect to the classical ``Type~I'' runaway typical of 
planetesimal populations around single stars.  This new 
type of growth termed ``Type~II'' runaway allows 
planet formation to occur in
binary-star systems with much tighter orbits than previously suggested
\citep{hep78,whit98}.  However, there is (at least) one caveat.  As already noted, 
the growth simulations described above and represented in Figure~\ref{typeII_runaway_fig2} 
did not include mutual perturbations of the planetesimals themselves. 
Crude attempts at including self-gravitating planetesimals
\citep{kortenkamp2000b} indicated that when the size distribution reaches $10^{24}$
to $10^{25}$ g the mutual perturbations are beginning to become
important, although they are still dominated by the size-dependent
phasing of secular perturbations.  This suggests that just as Type~II
runaway growth is getting underway accretion may either stall or perhaps transition to the classical
Type~I runaway growth.  To explore these possibilities we are working to modify the multi-zone planetesimal accretion code of \citet{weidenschilling1997} to include secular perturbations from massive companions as well as mutual perturbations from planetesimals and nebular gas drag \citep{kortenkamp2006}.

\section*{Acknowledgments}

S. Kortenkamp acknowledges support from NASA for some of this work under grants NNG04GP56G and NNG04GI14G.

{} 

\end{document}